# Generation of Quantum Vortex Electrons with Intense Laser Pulses

Zhigang Bu[1], Liangliang Ji[1,*], Xuesong Geng[1], Shiyu Liu[1,2], Shaohu Lei[1,2], Baifei Shen[3], Ruxin Li[1,4], and Zhizhan Xu[1]

[1]*State Key Laboratory of High Field Laser Physics, Shanghai Institute of Optics and Fine Mechanics (SIOM), Chinese Academy of Sciences (CAS), Shanghai 201800, China*
[2]*University of Chinese Academy of Sciences, Beijing 100049, China*
[3]*Department of Physics, Shanghai Normal University, Shanghai 200234, China*
[4]*Shanghai Tech University, 201210 Shanghai, China*

**Accelerating a free electron to high energy forms the basis for studying particle and nuclear physics. Here it is shown that wavefunction of such an energetic electron can be further manipulated with femtosecond intense lasers. During the scattering between a high-energy electron and a strong laser pulse, we find a regime where the enormous photon spin angular momenta can be efficiently transferred to the electron orbital angular momentum (OAM). The wavefunction of the scattered electron is twisted from its initial plane-wave state to quantum vortex state. Nonlinear quantum electrodynamics (QED) theory suggests that GeV-level electrons acquire average intrinsic OAM beyond $\langle l \rangle \sim 100\,\hbar$ at laser intensities of $10^{20}\,\text{W/cm}^2$ with linear scaling. These electrons emit gamma photons with double-peaked spectrum, which sets them apart from ordinary electrons. The findings demonstrate a proficient method for generating relativistic leptons with quantum vortex wavefunctions based on existing laser technology, thereby fostering a novel source for particle and nuclear physics.**

---

[*]e-mail: jill@siom.ac.cn

**INTRODUCTION**

High energy particles, such as relativistic electrons, play a crucial role in advancing our understanding of the subatomic world. These particles undergo acceleration to attain substantial momentum in free space and then collide with nuclei or other free particles. In many instances, their behavior is effectively described using plane-wave states, which are eigenstates of momentum and spin. Both features have been extensively utilized in the fields of particle and nuclear physics. In recent years, it has been found wavefunctions of not only photons but also charged particles can be reshaped, by spiral phase plates[1-4], fork gratings[3,5], or free electron lasing[6-9], to form coherent vortex states carrying intrinsic orbital angular momenta (OAM). Electrons with vortex wavefunctions (referred to as quantum vortex states, QVS) at keV level have been realized in transmission electron microscopes[10-13], enabling enhanced resolution and new insights into materials[14,15]. Yet these approaches are only limited for non-relativistic electrons with energies $\ll$ MeV. It remains unclear whether the wavefunctions of highly relativistic particles can be intentionally manipulated within existing accelerators or colliders, since their wavelengths are too small for any structures accessible with current technologies.

Here we show that intense laser pulses are efficient in transforming the wavefunctions of relativistic electrons. It relies on the scattering of electrons with multiple laser photons in a circularly-polarized (CP) pulse, i.e., the nonlinear Inverse Compton Scattering (NCS). The absorption of numerous laser photons by the electron not only leads to the transfer of momentum to the final particles but also angular momentum. It has been shown in NCS process that the emitted γ-photon could inherit the intrinsic OAM from the angular momenta (AM) of laser photons while its energy is blue-shifted[16-18]. However, the high-energy electrons involved in the interaction are considered unable to gain net intrinsic OAM[17,19-22].

We find a new regime where the electron gains significant portion of AM during NCS and carries intrinsic OAM. Our vortex scattering theory reveals that through radiation-reaction (RR) effect in the strong nonlinear regime, the average intrinsic OAM obtained by GeV-level electrons is proportional to the laser intensity, achieving beyond $100\,\hbar$ at laser intensities of $10^{20}$W/cm$^2$. With these lasers readily accessible in many facilities worldwide, the mechanism provides a new approach to manipulate the wavefunctions of highly relativistic electrons or positrons. It opens up new possibilities for investigating AM physics in nuclear systems[23,24] and creating exotic high-AM particle states in colliders relevant to high-energy physics, which are not possible with ordinary sources.

## RESULTS
### Converting non-vortex electron to QVS electron.

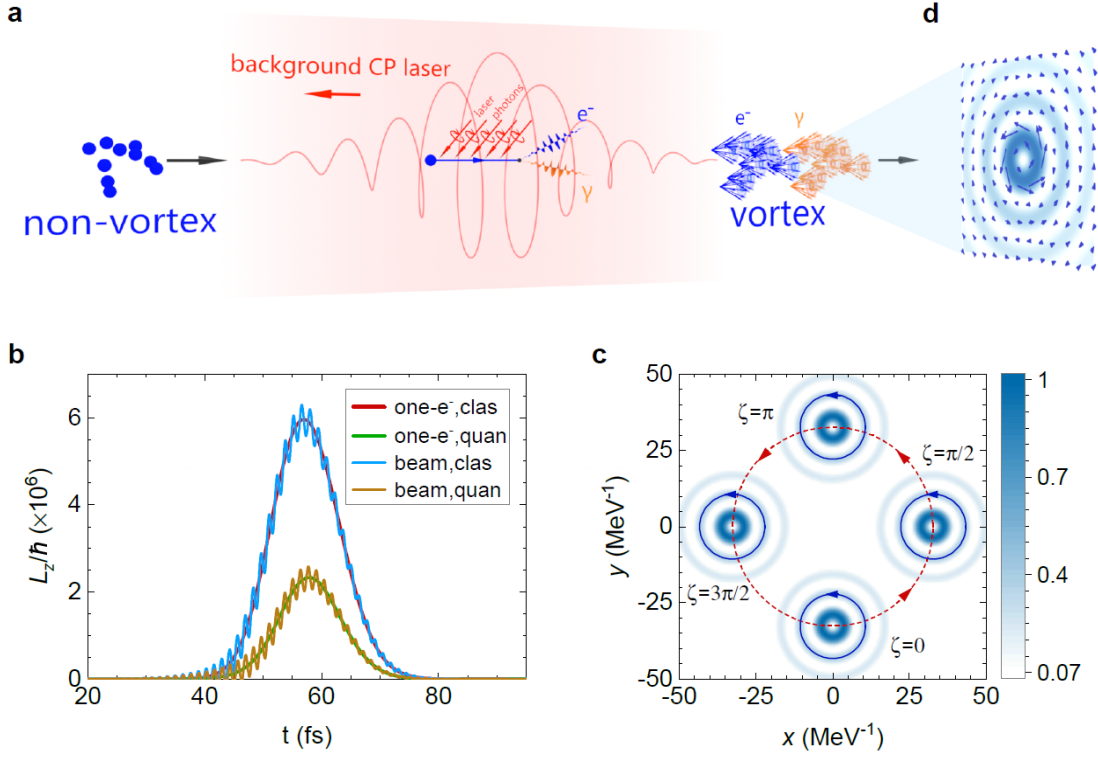

**Fig. 1 | Conversion from non-vortex electron to QVS electron in NCS process. a**, Schematic diagram of vortex NCS in CP laser fields, a high-energy electron absorbs $N$ laser photons and emits a QVS γ-photon, during this process it gains net intrinsic OAM and its wave function is twisted into vortex state. **b**, Time evolution of one-electron and electron beam OAMs when moving in a CP laser pulse with classical radiation and quantum radiation. **c**, Time evolution of the normalized probability density of Volkov-Bessel state in one laser period, red circle denotes the extrinsic circular motion driven by the CP laser field and blue circle denotes the intrinsic OAM of the vortex state. **d**, Transverse current density of QVS electron in CP laser field at propagation phase $\zeta = \pi$. The parameters of laser and electrons are $a_0=1$, $\lambda_0=1$, $l=1$, $s=1$, $\gamma_p=10^4$, the electron opening angle is defined by $\tan\theta_p = p_\perp/p_z$.

The concept is depicted in Fig. 1a by colliding GeV-level electrons with an intense CP laser pulse. In classical or semi-classical picture, a point-like electron in CP laser undergoes helical motion and acquires incoherent and extrinsic OAM[25-28]. However, this motion cannot be preserved after the interaction. As an example, Fig. 1b shows the time evolution of OAMs from a single electron or electron beam in the CP-laser-electron interaction. Here the electron radiation and induced reaction effect are considered with the classical Landau-Lifshitz model[29] and stochastic radiation model[30-32], respectively. In these cases, the OAM value is derived by counting $\mathbf{r} \times \mathbf{p}$ for each point-like particle, hence dependent on the choice of axis and referred to as extrinsic[33,34]. We see that OAMs in all cases increase in the first half of the laser pulse and then decline in the second. Eventually electrons do not acquire net extrinsic OAM as point-like particles after the interaction. It is also impossible to learn any information about the intrinsic properties of particle state in classical and semi-classical theory.

To find this out, we move to the quantum realm by describing particles with wavefunctions. In contrast to the widely used plane-wave state, the wavefunction here must include quantum numbers relevant to OAM, which can be characterized by vortex state in Bessel mode[35]. Further, the state is dressed in intense laser field during the interaction — often described by Volkov state[36]. Combining both features, we construct the quantum vortex state (QVS) of electron dressed in CP laser field $\psi^{s,l}_{p_\perp,p_z}(x)$, by coherent superposition of Volkov states $\psi^s_p(x)$, through helical Fourier spectrum (see in Methods). This wavefunction, known as Volkov-Bessel state, is depicted by electron energy and momenta $(E_p, p_\perp, p_z)$, intrinsic OAM number $l$ and SAM number $s$. It is not an eigenstate of any angular momentum projection due to the external field. Yet one can alculate the mean values of OAM projection onto the propagation z-axis, $\langle L_z \rangle = l + \Delta l$. The first term is the quantum number $l$, representing the value of intrinsic OAM; while the second one $\Delta l$ is determined by laser field strength and electron momentum, i.e., it is associated with the extrinsic OAM and spin-orbit interaction (SOI) dressed by the laser field. The differences can be seen clearly from the probability density $\rho^{s,l}(x)$ of electron state, its evolution in one laser period $[0, 2\pi]$ has been shown in Fig. 1c. At each phase, the concentric ring profile is well retained, corresponding to the intrinsic vortex structure of the electron state. Meanwhile, the entire density distribution rotates around a central axis for a cycle within one phase period. It is a typical motion of an electron moving in the CP laser field, generating extrinsic OAM $\Delta l$.

Using the QVS, we can see how the electron state is transformed into vortex state. The vortex structure is clearly shown by the transverse current density divided into intrinsic and extrinsic components, $\boldsymbol{j}^{s,l}_\perp(x) = \boldsymbol{j}^{s,l}_{\perp,int}(x) + \boldsymbol{j}^{s,l}_{\perp,ext}(x)$. Fig. 1d displays the intrinsic current vector at propagation phase $\zeta = \pi$, showing distinctive vortex structure. When electron leaves the laser field after the scattering, the extrinsically rotational motion (extrinsic current density) vanishes, while the intrinsic vortex structure is retained and simplified to the SOI of a free QVS electron[35]. In other words, the preservation of OAM manifests as the electron state becoming twisted into QVS.

Twist of the electron wavefunction is quantitatively analyzed in the framework of quantum electrodynamics (QED), where the scattering probabilities is obtained based on the Volkov-Bessel state. The multi-photon NCS is then 'simplified' to the decay of a non-vortex electron into a QVS photon and a QVS electron in the CP laser field (NV→V+V process). The AM-dependent differential scattering rate is derived from the S-matrix of the scattering, dP = $|S_{fi}|^2 p_\perp k_\perp dp_\perp dp_z dk_\perp dk_z$. Here we consider the laser pulse with frequency $\omega_0$ = 1.55eV, helicity $\lambda_0 = \pm 1$, intensity $I_0$ = 2.14×10$^{22}$ W/cm$^2$, normalized amplitude $a_0$ = |e|A$_0$/M = 100 (e and M are the charge and mass of electron, A$_0$ is the vector potential amplitude of the laser electric field) and duration L = 26 fs propagating along z-axis and colliding head-on with a relativistic electron. The electron is described by a wave packet of central energy $E_Q$ = 5.11 GeV (relativistic factor $\gamma_Q$ = 10$^4$). The opening angle of emitted QVS γ-photon defined by tan$\theta_k$ = $k_\perp/k_z$ is around $\theta_0$ = π-$a_0/\gamma_Q$ for $\gamma_Q \gg a_0$ [37]. Fig. 2a displays the emission rate of γ-photon after absorbing N laser photons at $\theta_0$, where the contributions of all AM modes have been summed. One sees that switching the wavefunctions of the final particles to QVS does not change the total emission rate. It is consistent with the one obtained in the ordinary non-vortex theory (red solid line). When absorbing N laser photons, the central energy of emitted γ-photon at $\theta_k$ can be approximated using the energy and momentum conservation condition,

$$\omega_c \approx \frac{N\omega_0 (E_Q - Q_z)}{(\tilde{E}_Q + N\omega_0) - \cos\theta_k (\tilde{Q}_z + N\omega_0)}, \qquad (1)$$

as illustrated in Fig. 2a, where $Q_z$ is electron central momentum along z-axis, $\tilde{E}_Q$ and $\tilde{Q}_z$ are electron quasi-energy and quasi-momentum in external CP laser field. In the limit $a_0 \ll 1$, $\theta_k \to \pi$ and N→1, it simplifies to the well-known linear form $\omega_c \to 4\gamma_Q^2 \omega_0$.

While the momenta of N laser photons are absorbed and converted to the emitted γ-photon following Eq. (1), the enormous SAMs are also necessarily transferred to final particles and turns their wavefunctions into vortex state. The connection between N and the intrinsic OAM number, l, of the scattered electron and the total angular momentum (TAM) number, j, of the γ-photon, follows the conservation law $N\lambda_0 = l+j+\Delta$, where Δ = 0, ±1 depends on whether the electron spin is flipped during the scattering. Under the parameter region considered here, the electron spin-flip rates are negligible, thus $N\lambda_0 \sim l+j$. We divide the emission rate in Fig. 2a into different AM channels and obtain the AM spectra in Fig. 2b with $N = 10^2, 10^3, 10^4$ and $5\times10^4$. The normalization by $N$ clearly shows the proportion of AM taken by final particles from the total SAM of absorbed laser photons. In the weakly nonlinear regime when $N$ is small $N < 10^4$, almost all the SAMs from laser photons are transferred to the $\gamma$-photon and the scattered electron does not acquire intrinsic OAM. It agrees with the cognition in linear Compton scatterings[16] where the electron efficiently transfers the SAMs of laser photons to the emitted photon while its own state remains minimally disturbed.

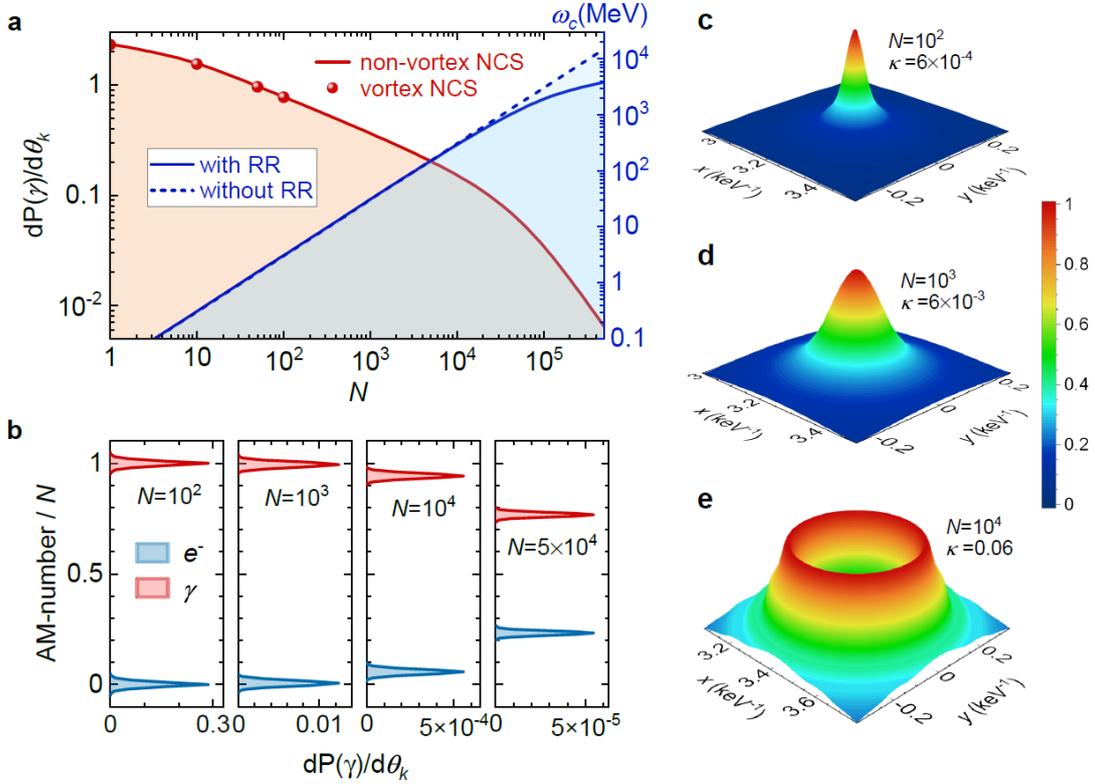

**Fig. 2 | Spectra of vortex NCS and transverse density of scattered electron. a,** Emission rate and central energy of $\gamma$-photons at opening angle $\theta_0$ versus $N$ calculated by the vortex scattering theory (red dots, summing over all AM modes) and non-vortex theory (red line). **b,** Intrinsic AM distribution of QVS electrons and $\gamma$-photons when absorbing $N = 10^2, 10^3, 10^4$ and $5\times10^4$ laser photons. **c-e,** Normalized density distributions of scattered electrons for $N = 10^2, 10^3$ and $10^4$.

The allocation shifts significantly when entering the highly nonlinear regime. As evidenced in Fig. 2b, when $N$ increases to be $>10^4$ the SAMs of laser photons are gradually transferred to the scattered electron, causing the spectrum center of $\gamma$-photon to shift from 1. Comparing the trend in Fig. 2a, one notices that the electron starts to gain significant OAM when the photon central energy deviates from the blue dashed line where the back reaction from $\gamma$-photon emission is omitted

(referred to as the non-Radiation Reaction case). We introduce a parameter $\kappa = \omega_c/E_Q$, defined as the ratio of the central energy of γ-photon to the incoming electron energy. This parameter not only governs the recoil effect on electron due to the emission of γ-photon but also determines the central intrinsic OAM number of scattered electron through the scaling

$$l_c = \kappa N, \qquad (2)$$

and the corresponding central TAM number of γ-photon is $j_c = N(1-\kappa)$. To see this, we rewrite Eq. (1) as $\omega_c \approx 2N\omega_0\gamma_Q/(a_0^2/\gamma_Q + 2N\omega_0/M)$. When $N\omega_0/M \ll a_0^2/\gamma_Q$, parameter $\kappa \ll 1$ and the photon central energy increases linearly with $N$, which is the non-RR regime. When $N$ is approaching $(a_0^2/\gamma_Q)(M/2\omega_0) \sim 10^5$, we have $\kappa \sim 1$ and the RR effect becomes apparent.

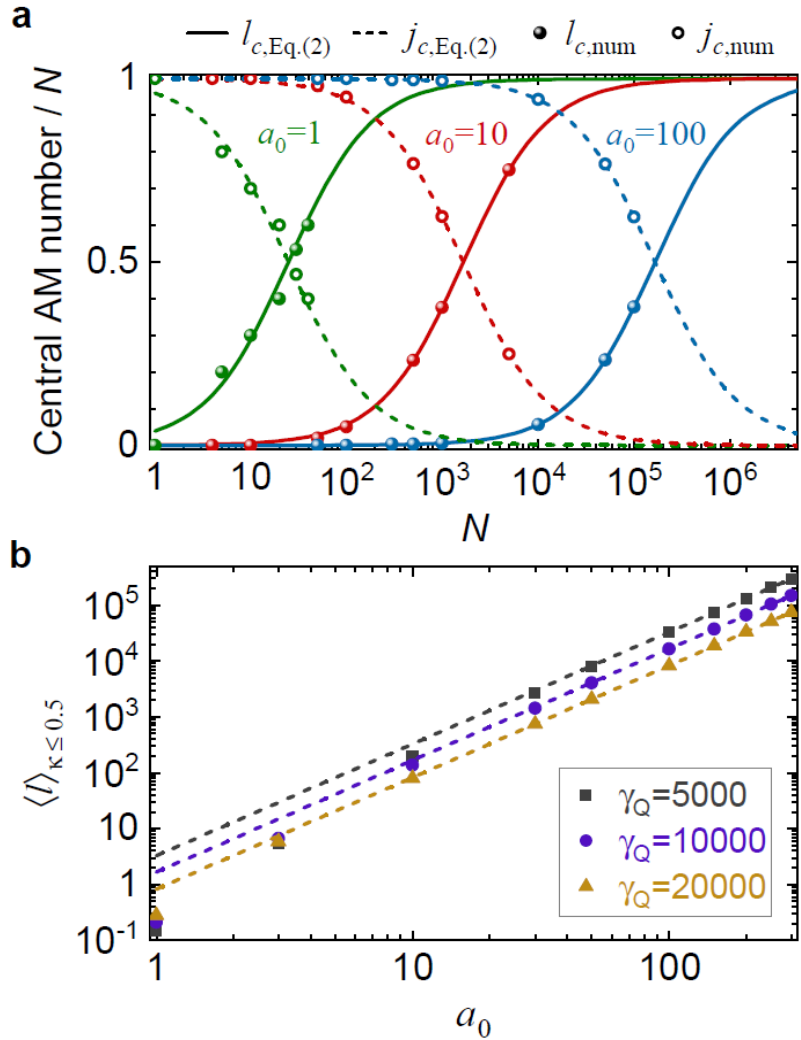

**Fig. 3 | Angular momentum spectra of scattered particles. a**, Central AMs of scattered electrons and γ-photons versus number of absorbed laser photons $N$. The laser amplitude is $a_0 = 1$, 10 and 100, the energy of incident electron is $\gamma_Q = 10^4$. **b**, Average value of scattered electron intrinsic OAM calculate in the range of $\kappa \leq 0.5$, $\langle l \rangle_{\kappa \leq 0.5}$. The dashed lines represent the $\propto a_0^2$ scaling.

In this regime the electron wavefunction is significantly twisted. In Fig. 2c to e, we plot the average transverse density of scattered electrons when absorbing $N = 10^2$, $10^3$ and $10^4$ laser photons. As $N$ increases and parameter $\kappa$ becomes non-negligible, the density profile undergoes a transition from a Gaussian-like distribution to a hollow structure, a typical sign coinciding with the quantum vortex structure and the electron OAM spectrum in Fig. 2b. Therefore, the production of QVS electrons in NCS is a novel effect triggered by RR in the quantum regime. This observation signifies the first acknowledgment of quantum RR affecting not just momentum but also angular momentum.

The distribution of the AM among the electron and γ-photon following Eq. (2) is confirmed by numerical results calculated from S-matrix, as shown in Fig. 3a. Here the central intrinsic AMs of electrons and γ-photons predicted by theory are in good agreement with the numerical calculation (dots) for a large range of laser intensities. Absorption of more laser photons causes an increase in the energy of emitted photon, thus $\kappa$ gradually becomes non-negligible and the electron wavefunction is twisted more significantly. A turning point appears at $\kappa = 0.5$ where RR is strong enough to cause the electron OAM to surpass the photon TAM. Beyond this point, the growth in photon TAM tends to saturate while the electron OAM gradually approaches $|l_c| \to N$.

The mechanism revealed here is rather efficient. We count the average value of intrinsic OAMs of the scattered electron in Fig. 3b as a function of the drive laser amplitude. Numerical results (dots) indicates that when laser amplitude $a_0 > 30$, the average intrinsic OAM scales approximately as $\langle l \rangle \propto a_0^2$ (dashed lines). At laser intensities of $10^{20}$ W/cm², one can achieve an angular momentum $\langle l \rangle \sim 100\,\hbar$. Notably, this level of intensity is readily attainable with 100TW-class laser systems, a capability already realized in numerous laser facilities. With state-of-the-art PW-class lasers, the average OAM approaches the order of $10^4\,\hbar$. There is no discernible significant limit for the average OAM value within the considered parameter regions.

**Scattering of QVS electron**

The scattered high-energy electrons with twisted wavefunction may undergo further scattering, invoking the V→V + V process. To resolve this, we describe the incoming QVS electron as a wave packet of Volkov-Bessel state with SAM and OAM numbers of $\tilde{s}$ and $\tilde{l}$. The electron central energy and momenta satisfy $E_Q = M\gamma_Q = \sqrt{Q_\perp^2 + Q_z^2 + M^2}$.

There is a significant difference in emission spectra between the vortex and non-vortex electrons, even in the weakly nonlinear regime. Considering the laser amplitude is $a_0 = 1$, the intrinsic OAM number of incoming QVS electron is $\tilde{l} = 1$, its central energy and transverse momentum are $\gamma_Q = 10^3$ and $Q_\perp = 2 \times 10^{-4} E_Q$. The energy spectrum of QVS γ-photons, where only one laser photon is absorbed, is depicted in Fig. 4a and shows a double-peaked distribution. In contrast, the NV→V+V process produces a single-peaked emission spectrum, as shown in Fig. 4b. The emission energy spectra at $\theta_k = 1$mrad are compared in Fig. 4c and d. The double-peaked structure of spectrum is attributed to the opening angles of incident QVS electron $\theta_q$ and emitted photon $\theta_k$. They form two angle configurations, $\alpha_1 = \pi - |\theta_k - \theta_q|$ and $\alpha_2 = \theta_k + \theta_q - \pi$, resulting in two central energies of emitted photons,

$$\omega_\pm \approx \frac{N\omega_0 (E_Q - Q_z)}{(\tilde{E}_Q + N\omega_0) + (\tilde{Q}_z + N\omega_0)\cos(\theta_k \mp \theta_q)}. \quad (3)$$

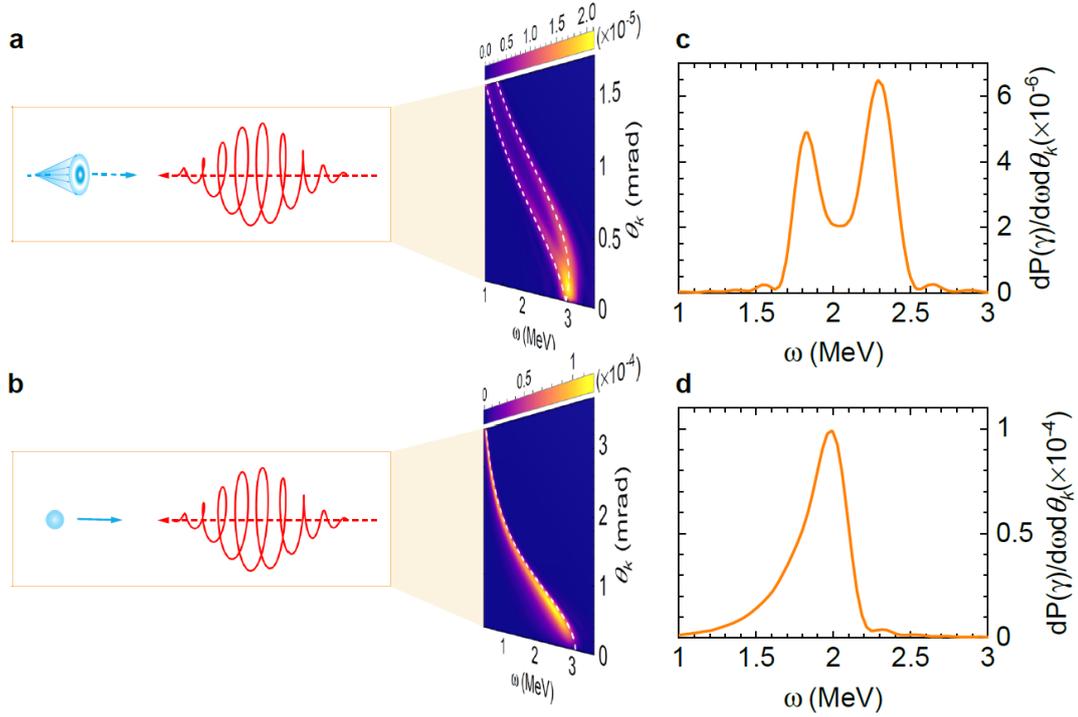

**Fig. 4 | Emission spectra of QVS and non-vortex electrons. a**, Emission spectrum of QVS electrons in CP laser field. **b**, Emission spectrum of non-vortex electrons in CP laser field. The white dashed lines represent the theoretically central energy of γ-photons. **c, d**, Energy spectra at emission opening angle $\theta_k = 1$mrad for incident QVS and non-vortex electrons.

Eq. (3) is depicted by the white dashed lines in Fig. 4a and b, they are in good agreement with the spectrum distribution. When $\theta_q \to \pi$, the incoming QVS electron degenerates into non-vortex electron, and Eq. (3) is simplified to Eq. (1). The double-peaked spectrum also serves as a distinct signal to differentiate the QVS electrons from the non-vortex ones, which can be achieved with current experimental techniques.

**CONCLUSIONS**

We demonstrated a new mechanism that leads to the twist of relativistic electron wavefunction using intense laser pulse. It is found that in the highly nonlinear regime, the RR effect dominates the transfer of intrinsic AM in vortex NCS dressed in CP laser field. We developed the first full-vortex scattering theory in the Furry picture of QED and found that the SAM of laser photons are mostly carried away by the emitted γ-photon in the weakly nonlinear regime. However, in the laser field with higher intensity, the triggering of RR effect drives the scattered electrons to gain higher portion of intrinsic AMs. The average intrinsic OAM acquired by scattered electrons scales approximately linear with the laser intensity. Compared to single-photon processes, intense laser pulse induces strong nonlinear effects (multiphoton absorption process), which significantly enhance the efficiency of QVS electron generation. The results show that when $a_0 > 10$, more than 90% of the scattered electrons gain intrinsic OAMs. After the interaction, these OAMs are effectively preserved and the electron wavefunctions are twisted into QVSs. Our work provides a viable scheme for manipulating the one-particle vortex state of high-energy electron with current laser technology, which paves the way for exploring intrinsic angular momentum and vortex effects in high-energy particle and nuclear physics.

**METHODS**

**Electron QVS in CP laser field (Volkov-Bessel state)**

The electron Volkov-Bessel state, $\psi_{p_\perp,p_z}^{s,l}(x)$, dressed in CP laser field is expressed as the superposition of Volkov states $\psi_p^s(x)$ [38,39],

$$\psi_{p_\perp,p_z}^{s,l}(x) = \int d\tilde{p}_z d\varphi_{\tilde{p}} d\tilde{p}_\perp \tilde{p}_\perp \psi_{\tilde{p}}^s(x) f_l(\tilde{p}), \quad (4)$$

where $f_l(\tilde{p}) = (1/\sqrt{2\pi}i^l\tilde{p}_\perp)\delta(\tilde{p}_\perp - p_\perp)\delta(\tilde{p}_z - p_z)e^{il\varphi_{\tilde{p}}}$ is the Fourier spectrum with helical phase. The probability density of Volkov-Bessel state is calculated as $\rho^{s,l}(x) = \psi_{p_\perp,p_z}^{s,l\dagger}\psi_{p_\perp,p_z}^{s,l}$. The transverse current density is given by $j_\perp^{s,l}(x) = \psi_{p_\perp,p_z}^{s,l\dagger}\boldsymbol{\alpha}_\perp\psi_{p_\perp,p_z}^{s,l}$.

The QVS of emitted photon, $A_\mu^{\lambda,j}(x)$, is constructed in a similar way using the plane-wave state[16,40], which is labeled by energy ω, momenta $k_\perp$ and $k_z$, helicity $\lambda = \pm 1$, and TAM projection $j$ along z-axis.

**S-matrix of the vortex NCS process**

According to the QED theory, the S-matrix of the NV→V+V process is

$$S_{fi} = -ie \int d^4x \psi_{p_\perp,p_z}^{s,l} \gamma^\mu A_\mu^{\lambda,j*} \Psi_Q^\sigma, \quad (5)$$

where $\Psi_Q^\sigma$ is the incoming non-vortex electron state and considered as a wave packet form of Volkov state, $\Psi_Q^\sigma(x) = \int d^3q \rho(q_\perp, q_z) \psi_q^\sigma(x)$, with central energy and momentum $E_Q$ and $Q_z$, $A_\mu^{\lambda,j}$ is the emitted QVS γ-photon state, $\psi_{p_\perp,p_z}^{s,l}$ is the scattered QVS electron state. Using the Gaussian wave packet for incoming electron, the S-matrix is derived,

$$S_{fi} \sim \int \frac{dq_z}{|q_z|} \delta(E_p + \omega - E_q - p_z - k_z + q_z)(e^{ig(q_z)} - 1)\rho_z(q_z) \xi^{s\dagger} \Xi_{k,p}^{j,l}(q_z) \xi^\sigma. \quad (6)$$

The delta function enforces the conservation of energy and longitudinal momentum. The phase factor $g(q_z) = n\pi(\tilde{E}_p + \omega - \tilde{E}_q + \tilde{p}_z + k_z - \tilde{q}_z)/\omega_0$, quantities with tildes are the electron quasi-energy and quasi-momentum in external CP laser field, satisfying $\tilde{E}_p^2 - p_\perp^2 - \tilde{p}_z^2 = \tilde{E}_Q^2 - \tilde{Q}_z^2 = M^2(1 + a_0^2)$. $\Xi_{k,p}^{j,l}$ is a 2×2 matrix, its four elements are linked to the spin polarization of incoming and outgoing electrons.

To calculate the V→V+V process, we replace the incoming electron state in Eq. (5) with the wave packet of Volkov-Bessel state.

In our calculations, we make use of the narrow wave packet approximation, which means the width of incident electron wave packet is much smaller than its central value.

**Parameter κ *versus* the QED parameter $\chi_0$**

Parameter $\kappa$ can be correlated with the Lorentz-invariant quantum parameter $\chi_0 = (E_0/E_{cr})(k_0 \cdot Q)/(\omega_0 M)$ as: $\chi_0 \approx (a_0^3/N)\kappa/(1-\kappa) = (a_0^3/N)(l_c/j_c)$. At the turning point $\kappa = 0.5$, we get $l_c = j_c$, and the quantum parameter $\chi_0 \approx a_0^3/N$. When the laser amplitude $a_0 = 1$ the number of absorbed laser photons $N = 25$ and the quantum parameter $\chi_0 \sim 0.04$ at the turning point so that the quantum effect are negligible. As the laser amplitude increases to 10 and 100, $N$ at the turning point increases to 1700 and $1.65\times10^5$ respectively, leading to larger values of $\chi_0$ (~ 0.59 and 6 respectively) and more significant quantum effects.

**Acknowledgements:** The authors would like to thank Dr. Karen Hatsagortsyan and Prof. Jianxing Li for valuable discussion and suggestion. This work is supported by National Science Foundation of China (Grant Nos. 12388102, 11935008), CAS Project for Young Scientists in Basic Research (Grant No. YSBR060), National Key R&D Program of China (Grant No. 2022YFE0204800).
**Author contributions:** Z.G.B. and L.L.J. proposed the research. Z.G.B. performed the theoretical and numerical analysis with suggestion from L.L.J.; X.S.G. helped to carry out part of the simulation. Z.G.B. and L.L.J. prepared the paper, with suggestion from X.S.G., S.Y.L., S.H.L, J.X.L., B.F.S., R.X.L. and Z.Z.X. **Competing interests:** The authors declare that they have no competing interests. **Data and materials availability:** All data needed to evaluate the conclusions of the paper are present in the paper.